# Experimenting with ChatGPT for Spreadsheet Formula Generation: Evidence of Risk in AI Generated Spreadsheets

Simon Thorne sthorne@cardiffmet.ac.uk

Cardiff Metropolitan University

**ABSTRACT**

Large Language Models (LLM) have become sophisticated enough that complex computer programs can be created through interpretation of plain English sentences and implemented in a variety of modern languages such as Python, Java Script, C++ and Spreadsheets. These tools are powerful and relatively accurate and therefore provide broad access to computer programming regardless of the background or knowledge of the individual using them. This paper presents a series of experiments with ChatGPT to explore the tool's ability to produce valid spreadsheet formulae and related computational outputs in situations where ChatGPT has to deduce, infer and problem solve the answer. The results show that in certain circumstances, ChatGPT can produce correct spreadsheet formulae with correct reasoning, deduction and inference. However, when information is limited, uncertain or the problem is too complex, the accuracy of ChatGPT breaks down as does its ability to reason, infer and deduce. This can also result in false statements and "hallucinations" that all subvert the process of creating spreadsheet formulae.

## 1.0 INTRODUCTION

### 1.1 Large Language Models

Large Language Models (LLM) such as ChatGPT or Google Baird are deep learning neural networks trained on vast corpora of human generated language (Jiang, et al., 2020). Over millions of iterations, the neural network builds a probability-based language model of different words it encounters in the corpus. This allows the algorithm to build a detailed probabilistic network of what words relate to one another, how they are used and what they mean in the context of similar terms. When a user inputs a prompt, the text is analysed, the most probable intent or context of the prompt is settled upon, and a corresponding response is provided. ChatGPT optimises 167 billion parameters, making it the largest and in theory the most capable LLM at present.

LLMs can also be leveraged for code generation in a variety of languages including spreadsheet formulae (Chen, et al., 2021). This is based on the same approach for natural language generation, a probabilistic output based on the most likely response to the prompt input by the user. The user inputs plain text prompts that describe and specify what the user wants the code to compute and ChatGPT provides the most likely solution based on the prompt.

### 1.2 Producing spreadsheet formulae in ChatGPT

Using plain English descriptions, it is possible to create spreadsheet formulae with ChatGPT. If used correctly, this is a powerful tool that could allow those who cannot wield the syntax of Excel or other spreadsheet languages to use the AI generated versions.

For instance, consider a prompt for creating a grading formula for student work at a university:

> *The grade is based on an average of two cells, if the average is 70 or greater then award a 1st, if the average is equal to 60 and less than than 69 then award a 2:1, if the average is equal to 50 and less than 59 award a 2:2, if the average is equal to 40 and less than 49 then award a third, otherwise award a fail.*

The above text prompt produces the following accurate spreadsheet formula:

=IF(AVERAGE(A1,B1)>=70,"1st", IF(AVERAGE(A1,B1)>=60, "2:1", IF(AVERAGE(A1, B1)>=50, "2:2", IF(AVERAGE(A1, B1)>=40, "3rd", "Fail"))))

However, the quality of the output is directly related to the quality of the input and the more specific the prompt, the more accurate the formulae that ChatGPT creates. If some details are omitted or not specified, the algorithm will omit and take the text literally.

Hence ChatGPT is a top-down approach to coding, one has to understand and describe all parameters of the problem in order to generate valid and useful code. Based on these observations a number of experiments were devised.

**1.3 Experiments with ChatGPT**

The overall aim of these experiments is to determine how accurate ChatGPT is at producing valid spreadsheet formulae and solving "computational" problems with uncertainty, inference and deduction.

A number of research questions and hypotheses are identified:

**Research Question 1:** How does ChatGPT perform code generation when it is required to solve an incompletely described problem?

**Research Question 2:** What underlying knowledge and competence does ChatGPT have in logic, deduction and inference?

**Hypothesis 1:** The accuracy of solutions provided by ChatGPT will vary depending on the amount of uncertainty in the problem description.

**Hypothesis 2:** ChatGPT has a limited ability to reason with logic, deduction and inference and this limited ability will be obvious when solving more complex logic problems

**1.4 Experimental tasks**

There are three separate experiments testing the accuracy of ChatGPT. These tasks will use different types of problems to test different aspects of the technology. All experiments are conducted with ChatGPT 4

**1.4.1 Experiment 1: problem solving with uncertainty**

The experiment task for problem solving is the wall task (Rakovic, et al., 2019; Teo & Tan, 1997; Irons, 2003; Panko & Halverson, 2001; Panko & Sprauge, 1998). The wall task is used as a standard way of illustrating how a relatively simple spreadsheet task can generate an array of errors and has been used many times in spreadsheet error research to demonstrate how easy it is to make mistakes and explore error rates in creating spreadsheets and is considered a benchmark in spreadsheet error creation. Whilst the wall task may be considered 'simple' in the world of spreadsheet modelling and error, the task may well be complex for a LLM and is unlikely to be included in the corpus of material ChatGPT is trained on. This means that ChatGPT will be providing answers without any existing information to fall back on and hence the output is going to be as extreme and as "naked" as possible. Depending on the answers provided by ChatGPT, a series of other prompts will be passed to clarify the details of the problem if needed. This in theory makes solving the problem *easier* for ChatGPT.

### 1.4.2 Experiment 2: logic problems

Logical deduction problems starting at the easier end of the spectrum and increasing in difficulty will be passed to ChatGPT. The easier puzzles are those that contain all the information needed to come to right conclusion with minimal inference or deduction. The medium difficulty puzzles are those that contain most of the information needed to come to the right conclusion but require logical deduction and inference. The hard level puzzles provide the bare minimum information needed and need significant logical deduction and inference. The puzzles are sourced from Goodman (2023) and all puzzles are descriptive word-based puzzles.

### 1.4.3 Experiment 3: Boolean Logic Problems

In this final stage of the experiment, Boolean logic questions will be posed to ChatGPT that are drawn from a first year Bachelor of Science course in Computer Science in Cardiff School of Technologies, Cardiff Metropolitan University. These questions probe logical equivalences in Boolean logic achieved by combing of standard and negated logic gates with negated input.

## 2.0 RESULTS

The follow section will discuss the results obtained from the multiple phases of the three experiments conducted.

### 2.1 Experiment 1 Phase 1 Standard Wall Task

In this initial phase, ChatGPT is given the wall task without any modification. This task is usually given to students to complete as a means to illustrate how even simple spreadsheets can yield mistakes and that these mistakes can be difficult to notice and correct.

If we give ChatGPT the unedited wall task, the result contains many mistakes. Before the prompt was entered, ChatGPT was told it was going to act as an Excel programmer. The prompt given was:

> *You are to build a spreadsheet model to help you create a bid to build the wall around the garden.*
>
> *You will offer two options, Welsh granite or Brick.*
>
> *The wall will be built by a team of two people. The team will work three eight hour days to build either type of wall.*
>
> *The wall will be 20 feet long 6 feet tall and 2 feet thick.*
>
> *Wages are £5 per hour per person. You will need to add 20% to the wages to cover incentive bonuses.*
>
> *Welsh granite costs £5 per cubic foot and brick costs £4 per cubic foot.*
>
> *Your bid must add a 30% profit margin to your expected cost.*

The response was:

> =IF(A1="Welsh Granite", (20*6*2) *5 *2 *3 *1.2 *1.3, IF(A1="Brick", (20*6*2) *4 *2 *3 *1.2 *1.3))

This output contains several problems and provides an incorrect result. The formulae will be broken down into its identifiable parts, examined for accuracy and how the placement of the text in the prompt influences the solution.

The first condition in the opening IF statement and the *value if false* which opens another IF statement, =*IF (A1="Welsh Granite"* and *IF(A1="Brick"* , hinges on a text value in A1, either "Welsh Granite" or "Brick".

The second part calculates the volume of the wall appearing as a calculation in brackets, *(20*6*2)*. This calculation is correct and accurately reflects the instructions given but omits the material cost calculation which would appear to be a logical inclusion at this stage in the calculation.

The third part calculates the wage cost, some of the material costs and the profit margin, *\*5 \*2 \*3 \*1.2\*1.3*. The first thing to note that unlike the volume calculation, the wage, material and profit margin costs are not in brackets. Secondly, what is omitted from this calculation is the 8 hours a day and £5 per hour wage parameters. The *\*5* relates to the Granite material costs (an additional query was passed to ChatGPT changing the £5 per hour to £6 per hour to test which "5" is being cited in this formula), the *\*2* relates to the number of workers needed, *\*3* relates to the three working days needed to complete the work and the *\*1.2* relates to the added 20% incentive bonus for the workers. The *\*1.3* relates to the 30% profit margin that should be added to the overall cost of the bid.

The lack of brackets mean that the calculation is completely inaccurate, there should be brackets around the whole statement (5 * 2 * 3) * 1.2))*1.3. Even this would not fix the problem though since the 8 hours a day and the £5 per hour aspects of the wage calculation are omitted from the answer, the logic of including the Granite cost per square foot in this part of the calculation is unclear. Incorporating the omitted elements would also mean that the calculation would have to be approached differently, a more logical approach would be to calculate the volume and material costs and then add those to the wage costs and then calculate the 30% profit margin.

### 2.2 Experiment 1 Phase 2

It is interesting that the solution offered by ChatGPT is an attempt to answer the whole problem in one cell. In order to see if the complexity of fitting everything in one cell is a limiting factor, another prompt was attempted adding the following sentence at the end of the problem description:

> *You should produce separate formulae for brick or granite costs, wage costs and profit margin calculations.*

The thinking behind this is that perhaps a valid solution can be arrived at if the complexity of the formulae was reduced. ChatGPT output the following:

> *Brick/Granite Costs: IF(A1="Welsh Granite", (20*6*2)*5, IF(A1="Brick", (20*6*2)*4))*
> *Wage Costs: (20*6*2)*5*2*3*1.2*
> *Profit Margin: (20*6*2)*5*2*3*1.2*1.3*

Here ChatGPT has produced some correct and incorrect calculations. The volume calculation is correct as it was in the previous example. The materials costs are provided correctly, this is a change to the previous solution offered, perhaps because the prompt provided to ChatGPT explicitly asked for the brick and granite costs. However, both the wage costs and the profit margin are incorrect. The wage cost still omits the 8 hour day calculation and both the wage and profit margin calculations do not use any brackets, other than the volume, to ensure the order of calculation is correct.

### 2.3 Experiment 1 Phase 3

A modified prompt was passed to ChatGPT which contained the problem statement as already stated but included the following addition.

> *The wages should be calculated on the number of hours worked in three days at £5 per hour with a 20% bonus added to the result.*

It was hoped that by explicitly stating the correct method for calculating wages that ChatGPT's response would be more accurate. ChatGPT produced the following:

> *Material Costs: IF(A1="Welsh Granite",(20\*6\*2)\*5, IF(A1="Brick",(20\*6\*2)\*4))*
> *Wage Costs: (2\*3\*8)\*5\*1.2*
> *Profit Margin: (20\*6\*2)\*5\*2\*3\*8\*5\*1.2\*1.3*

The solution provided shows a correct wage cost calculation with no omissions and correct use of brackets. However, the profit margin calculation, which brings together the wage, material, bonus and profit margin costs does not use brackets to ensure the calculation is executed in the proper order. The wage cost in the profit margin calculation is different to the one provided under wage costs, the difference being missing brackets.

ChatGPT was then asked to provide the solution as one formula, the response was as follows:

> *=IF(A1="Welsh Granite", ((20\*6\*2) \*5 \*2 \*3 \*8 \*5 \*1.2 \*1.3), IF(A1="Brick", ((20\*6\*2) \*4 \*2 \*3 \*8 \*5 \*1.2 \*1.3)))*

Clearly from the above outcome, the original omission made by ChatGPT is corrected but the calculation is not correct because of the incorrect order of computation through a lack of brackets. The correct method using brackets for calculating the wage cost is abandoned in favour of a solution that closely resembles the original solution provided by ChatGPT.

### 2.4 Experiment 2 Phase 1

In this initial phase of the experiment, *easy* logic problems are posed to ChatGPT. These easy problems contain all of the information needed in the question and require no or little logical deduction or inference.

The prompt posed was:

> *Six children (Leisha, Benito, Delia, Charlotte, Weldon, and Zina) were measured yesterday. Weldon is taller than Delia but shorter than Zina. Leisha is taller than Benito but shorter than Delia and Weldon. Benito is not the shortest. Can you put them in order of height from tallest to shortest?*

The answer provided by ChatGPT was:

> *Zina, Weldon, Delia, Leisha, Benito, Charlotte*
>
> *Zina is the tallest, as Weldon is taller than Delia but shorter than Zina. Weldon is next, followed by Delia. Leisha is taller than Benito but shorter than Delia and*

> *Weldon, so she is next. Benito is not the shortest, so he must be next. Charlotte is the shortest, as she is the only one left. Therefore, the order from tallest to shortest is Zina, Weldon, Delia, Leisha, Benito, Charlotte.*

As can be seen from the answer, ChatGPT has correctly solved the problem and cited the correct deduction from the text. This success is likely due to the minimal deduction needed and the completeness of the information in the prompt.

**2.5 Experiment 2 Phase 2**

In this phase of the experiment, more complex logic puzzles were presented to ChatGPT and the outcome observed.

The first prompt given to ChatGPT:

> *Each of five neighbourhood dogs (Saber, Ginger, Nutmeg, Pepper, and Bear) is enjoying one of the following activities: getting its ears scratched, playing catch, taking a nap, burying a chew toy, and going for a walk.*
>
> *Pepper is either playing catch or burying a chew toy. Neither Ginger nor Saber nor Bear is on a walk. One of the dogs named after a spice is getting its ears scratched. A dog not named for a spice is playing catch. Bear is getting some exercise. Can you figure out what each dog is doing?*

The answer and reasoning provided by ChatGPT to the prompt:

> *Saber: Taking a nap Ginger: Burying a chew toy Nutmeg: Getting its ears scratched Pepper: Playing catch Bear: Going for a walk*
>
> *This is based on the clues given in the question. Pepper is either playing catch or burying a chew toy, so one of those two activities must be what it is doing. Neither Ginger, Saber, nor Bear is on a walk, so none of them are doing that activity. One of the dogs named after a spice is getting its ears scratched, so that must be Nutmeg. A dog not named for a spice is playing catch, so that must be Pepper. Finally, Bear is getting some exercise, so it must be going for a walk.*

The answer provided is inaccurate and the reasoning provided contains several contradictions. The correct answer to the problem is as follows: Saber is asleep, Ginger is getting her ears scratched, Nutmeg is on a walk, Pepper is burying a chew toy and Bear is playing catch.

This conclusion is reached through a series of deductions, the first two clues reveal which dog is on a walk, since either Pepper is playing catch or burying a chew toy AND neither Ginger, Saber or Bear is on a walk, that only leaves Nutmeg left.

The second step is given in the third clue, one of the dogs named after a spice is getting its ear scratched. Since we know Nutmeg is on a walk and Pepper is either burying a chew toy or playing catch, it can only be Ginger since neither Bear nor Saber are named after a spice.

The third step is given in the fourth, fifth and sixth clues, a dog not named after a spice, either Bear or Saber, is playing catch. The fourth clue reveals it must be Bear since Bear is getting some exercise but according to the sixth clue is "not on a walk". This means Bear must be playing catch since that is the only activity left that counts as exercise. This also means that Pepper must be burying a chew toy since she is not playing catch and that Saber must be asleep since this is the only activity not accounted for at this stage.

The answer provided by ChatGPT is mostly wrong and the reasoning cited is contradictory in places. The only correct assertion in the answer is that Saber is asleep. All of the other conclusions are incorrect and there is little evidence of logical reasoning. A few examples of this will now be highlighted:

When deducing which dog is getting its ear scratched, ChatGPT states the following:

> *One of the dogs named after a spice is getting its ears scratched, so that must be Nutmeg*

At this stage, there are several candidates who could fit the description in the problem but Nutmeg is not one of them. Nutmeg is identified in the first clue as going on a walk since it cannot be Pepper and through negation it cannot be Saber, Bear or Ginger.

When deducing what activity Pepper is involved in, ChatGPT states:

> *A dog not named for a spice is playing catch, so that must be Pepper*

Here ChatGPT doesn't seem to recognise that Pepper is named after a spice although it could possibly think that Pepper is a vegetable, in either case the answer provided is incorrect and Pepper is a commonly known spice It is unclear whether the *not* is a recognised logical operator here since the response ignores the implication of *not*.

When deducing what activity Bear is involved in, ChatGPT states the following:

> *Finally, Bear is getting some exercise, so it must be going for a walk.*

However, this contradicts the second clue that states:

> *Neither Ginger nor Saber nor Bear is on a walk*

So it is unclear why ChatGPT would come to the conclusion that Bear is on a walk since it is explicitly stated otherwise. It would appear that *Neither* and *nor* are not understood logical operators, this may explain why ChatGPT came to the incorrect conclusion about Bear being on a walk.

### 2.6 Experiment 2 Phase 3

The final prompt passed to ChatGPT is as follows:

> *The Reds, the Grays, the Blues, and the Greens have a round-robin tournament. Each team plays each other team once, for a total of six games. The Greens won more games than the Blues. The Grays lost more games than the Blues. The Reds tied the Greens. (This was the only tie in the tournament.) Who won the game between the Reds and the Blues?*

The correct response and reasoning is as follows: The Reds won the game between the Reds and the Blues. The Greens tied one game, so they won a maximum of two games. The Greens won more than the Blues, so the Blues won a maximum of one game. Since the Greys lost more times than the Blues, the Greys must have lost all three matches. The match that the Blues won must have been against the Greys, so the Blues lost the match against the Reds.

The response from ChatGPT is as follows:

> *The Reds and the Blues did not play each other, so there is no winner for that game.*

> *The Reds and the Blues did not play each other in the round-robin tournament, so there is no winner for that game. The clues given in the question only provide information about the results of the other five games. The Greens won more games than the Blues, the Grays lost more games than the Blues, and the Reds tied the Greens. Therefore, there is no winner for the game between the Reds and the Blues.*

The response from ChatGPT is a little surprising but incorrect. The conclusion that the Reds and Blues did not play each other is contradicted in the prompt by two statements, the first is the definition of a 'Round Robin' and the second is that there were six games played.

> *The Reds, the Grays, the Blues, and the Greens have a round-robin tournament. Each team plays each other team once, for a total of six games*

This total failure to compute the answer is no doubt based on the lack of information available to ChatGPT in the prompt. For instance, in the response to the prompt, ChatGPT complains that there is no information about the Red versus Blue game, *The clues given in the question only provide information about the results of the other five games* and therefore incorrectly concludes that there was no game between Red and Blue.

### 2.7 Experiment 3 Phase 1

In this phase, ChatGPT was asked to provide all of the standard and negated logic gate truth tables. The response obtained from ChatGPT was a complete list of correct truth tables covering OR, AND, XOR, NOT, NOR, NAND and XNOR.

ChatGPT was also asked what negated input in truth tables was, ChatGPT provided the following correct answer:

> *Negated input in truth tables is when the input is reversed. For example, if the input is A and B, then the negated input would be NOT A and NOT B. This means that the output of the truth table will be the opposite of what it would be with the original input.*

This first phase was completed to establish if ChatGPT could demonstrate logic gates, truth tables and negated input needed in phases 2 and 3.

### 2.8 Experiment 3 Phase 2

Given the responses from ChatGPT in experiment 3 phase 1, a correct explanation of negated input and the correct reproduction of logic gate truth tables, the following prompt was posed:

> *Does a negated input NOT gate produces an output that is the same as the original input?*

ChatGPT responded:

> *No, a negated input NOT gate produces an output that is the opposite of the original input.*

This answer is incorrect, the correct response is yes, a negated input not gate produces an output that is the same as the input:

| -A | A | NOT |
|----|---|-----|
| 0  | 1 | 0   |
| 1  | 0 | 1   |

**2.9 Experiment 3 Phase 3**

Given the responses from ChatGPT in experiment 3 phase 1, a correct explanation of negated input and the correct reproduction of logic gate truth tables, the following prompt was posed:

> *Does a negated 2-input OR gate produce an output that is the same as a 2-input NAND gate?*

The correct answer to this question is yes, these two combinations of OR, AND and NOT are equivalents:

| -A | -B | OR | A | B | NAND |
|---|---|---|---|---|---|
| 0 | 1 | 1 | 1 | 0 | 1 |
| 1 | 0 | 1 | 0 | 1 | 1 |
| 0 | 0 | 0 | 1 | 1 | 0 |
| 1 | 1 | 1 | 0 | 0 | 1 |

ChatGPT provided the following response:

> *No, a negated 2-input OR gate does not produce an output that is the same as a 2-input NAND gate. A 2-input OR gate produces an output that is the logical sum of its two inputs (i.e., if either one or both of the inputs is high, the output will be high). On the other hand, a 2-input NAND gate produces an output that is the logical product of its two inputs (i.e., both inputs must be high in order for the output to be high). Therefore, the output of a negated 2-input OR gate is not the same as the output of a 2-input NAND gate.*

The response is incorrect, the reasoning cited is partially correct and partially incorrect. The first assertion made by ChatGPT is correct but subsequent assertions are incorrect, the responses seem to ignore the implication of NOT.

The definition of OR is correct: *A 2-input OR gate produces an output that is the logical sum of its two inputs (i.e., if either one or both of the inputs is high, the output will be high)*. The error comes when discussing NAND, *a 2-input NAND gate produces an output that is the logical product of its two inputs (i.e., both inputs must be high in order for the output to be high)*. Here ChatGPT has given the definition of an AND gate and has ignored the NOT implication.

**3.0 CONCLUSIONS**

The experiments have provided some interesting insight into how capable ChatGPT is in different scenarios. This section will now address the research questions and hypotheses identified at the start of the paper.

**3.1 Research Question and Hypothesis 1**

Research question 1 and hypotheses 1 were tested in experiment 1, phases 1, 2 and 3 and in experiment 2, phases 1, 2 and 3. In the first set of experiments, the wall problem was posed to ChatGPT.

In experiment 1 phase 1, ChatGPT made some significant errors in the solution it offered. It made errors of omission when calculating the wage costs and it made BODMAS errors in the final calculation which meant the answer was completely wrong. The conclusion reached was that the problem description was too complex, phases 2 and 3 attempted to reduce the uncertainty and problem solving needed.

In experiment 1 phase 2, the prompt asked for formulae for all of the sub tasks to be expressed separately, rather than in one long formulae which ChatGPT naturally defaults to. This resulted in some improvement, the material costs were correctly calculated but the wage cost still omitted relevant information and BODMAS was still an issue.

In experiment 1 phase 3 explicit instructions on exactly how to calculate wage costs were included in the prompt passed to ChatGPT. The results obtained showed some improvement, the wage cost was correctly calculated with the correct use of BODMAS as a separate calculation. However, when ChatGPT was asked to use the solution in one formulae, it kept the correct method for calculating wages but abandoned the correct use of BODMAS making the result totally inaccurate.

The BODMAS errors in all three phases of experiment 1 are consistent with other observations on the mathematical failures of ChatGPT although BODMAS is not explicitly addressed (Frieder, et al., 2023; Borji, 2023).

In the second series of experiments ChatGPT was passed a series of logic puzzles, some of which contained all of the information needed to compute the correct answer, some of which needed deduction and inference to solve them correctly.

In experiment 2 phase 1, a simple sorting logic puzzle with no uncertainty or inference was posed to ChatGPT. The result provided was completely correct.

In experiment 2 phase 2, a relatively more complex logic puzzle than phase 1 was used which required some inference. The response generated was incorrect and contained contradictions and errors which seem to stem from the use of negation in the description of the problem. The response contained only one correct assertion and ChatGPT would seems to ignore the use of negation, specifically the use of the terms *neither* and *nor*.

In experiment 2 phase 3, a complex logic puzzle was posed to ChatGPT that needed significant amounts of inference and deduction to compute the right answer. ChatGPT gave a slightly unexpected answer to the question, the answer provided contested the assertion made in the prompt given to it on the basis that no information was provided about the specific prompt posed to ChatGPT.

In terms of answer research question 1, the evidence generated from the experiments suggests that ChatGPT has very limited ability to generate valid code where it is required to problem solve the question. This is mostly evident in experiment 1 where the initial answer provided to the unedited version of the wall task contained some correct but mostly incorrect responses to the prompt posed. In phases 2 and 3, the prompt was simplified in an effort to reduce the problem solving needed and make the requirements for the computation more explicit. This simplification did improve the accuracy of the responses but it did not result in a completely correct solution.

This is also supported by the outcome of experiment 2, where logic deduction puzzles were posed to ChatGPT which contained varying levels of uncertainty. Uncertainty, deduction and inference all negatively impact the ability of ChatGPT to respond correctly as shown in the results to phases 2 and 3 of experiment 2. According to Bang et al (2023), ChatGPT shows limited ability in inductive reasoning and in mathematical reasoning in general, a finding also echoed other researchers (Davis, 2023; Frieder, et al., 2023)

Hypothesis 1 is therefore upheld, the amount of uncertainty in the prompt does limit the accuracy of the solutions provided, in the experiments conducted, uncertainty comes in the form of a lack of complete information and the amount of deduction and inference needed to compute the right answer. In scenarios where there is no deduction or inference and complete information about the purpose of the task, the performance of ChatGPT is efficient and accurate.

### 3.2 Research Question and Hypothesis 2

Research question and hypothesis 2 was designed to probe why ChatGPT struggles with logical deduction, inference and uncertainty. The research question and hypothesis are answered by experiments 2 and 3.

Experiment 2 showed that as the prompt demanded more uncertainty, logical deduction and inference, the accuracy and performance of ChatGPT diminished.

Other authors have experimented with logical reasoning and ChatGPT, posing similar logic puzzles to those contained in this paper and noting the outcome (Borji, 2023). The results broadly agree with the outcome of the experiments in this paper, ChatGPT is unable to deduce the correct answers to these puzzles, no real reasoning is given for this however.

The results from experiment 3 show that despite ChatGPT having all of the information needed to compute the queries put to it, it was unable to come up with a correct assertion. This is evidenced by the correct definition of negated input and production of logic gates and truth tables for common logic operators and not combinations such as AND, OR, XOR, NAND, NOR, XNOR and NOT. However, when logic equivalence questions were posed to it, it was unable to arrive at the right answer and provided incorrect reasoning. Although many parts of the reasoning were incorrect, ChatGPT seems to have a particular issue with negation. Some of the reasoning provided would have been true if the prompt didn't included gates that are negated such as NAND or NOR.

The answer to research question 2 is that ChatGPT has only shallow "knowledge" of logic operators and although it can correctly cite the function and composition of logic gates, it is unable to apply this knowledge in the queries posed to it. Further, ChatGPT doesn't really "know" anything, whilst it may be able to provide the correct information given a prompt passed to it, ChatGPT doesn't "understand" the information it is passing to the user, **it is simply the most probable answer given the input prompt.** Indeed, all of ChatGPTs "knowledge" is the same, the answers provided are merely labels that have a high probability of fulfilling the intent of the prompt.

Some studies (Wang, et al., 2023) probe the use of Boolean logic for literature searching and achieve comparable results to other tools, however they do not specifically test ChatGPTs ability to reason with Boolean logic. So whilst there a consensus that ChatGPT struggles with written logic puzzles, this paper is novel in presenting results obtained from Boolean logic expressions hence there is no secondary work to compare these results with.

Hypothesis 2 is upheld, ChatGPT has only shallow, surface knowledge of logical reasoning and as such it's ability to provide logic deduction is almost entirely based on the availability of information in the prompt, if some information is missing, ChatGPT is unable to fall back on logical reasoning to solve the problem and instead produces incorrect solutions where it is not explicitly told how to reason (Bang, et al., 2023; Borji, 2023; Frieder, et al., 2023).

### 3.3 Other observations on ChatGPT

During the development and execution of the experiments, a number of interesting observations were made about the behaviour of ChatGPT. This section will explore some of these observations and what can be learnt from it.

### 3.3.1 Inconsistent use of BODMAS

When executing the wall task experiment, without exception, ChatGPT would always use the correct formulae for calculating volume including the correct use of BODMAS. This is in contrast to other solutions provided which contained incorrect use of BODMAS, resulting in incorrect calculations. At first this was puzzling but it was realised that it was consistently able to provide a correct calculation for volume because **it is highly likely that the formulae for volume is explicitly in the large language corpus on which ChatGPT is trained**. So it's able to cite and calculate volume correctly because it has already learnt the explicit way in which it should be calculated. However, when it is required to apply BODMAS in a situation not explicitly covered in the training of the model, it is unable to use the same principles correctly, inconsistencies in logic and mathematical reasoning are cited by many studies (Borji, 2023; Bang, et al., 2023; Davis, 2023; Frieder, et al., 2023).

The same may also be said of the correct answers provided around negated input and various logic gates. It is highly likely that **these gates form part of the large language corpus** and hence it can reproduce them accurately. However, when it faces novel situations where it has to apply that knowledge, it is unable to complete this task since it is not explicitly covered in its training phase. If the exact questions used in these experiments had been part of the training corpus, then it's likely that ChatGPT would have provided the correct answers.

### 3.3.2 False statements

The manner in which the prompt is worded makes a great difference to how ChatGPT responds. When posing logic questions in experiment 3, the questions were lifted from a class on Boolean logic. The unedited wording of the questions started "Show that" a negated two input OR gate produces the same output as 2 input NAND gate". Specifically, using "show that" phrase induced ChatGPT into inventing false proofs and reasoning to "show that" the two outputs are the same. This was confusing when initially encountered because the correct answer to the question is yes, they are equivalents but in examining the reasoning behind the answer generated, it becomes clear that the reasoning provided is false.

To test this idea further, a prompt was posed that said "Show that a 2 input AND gate produces the same output as a 2 input OR gate". This is obviously impossible since the outputs of these two gates are different. The reasoning in response to this impossible question is false and contradicts the accurate truth table provided by ChatGPT:

| A | B | A AND B | A OR B |
|---|---|---------|--------|
| T | T | T       | T      |
| T | F | F       | T      |
| F | T | F       | T      |
| F | F | F       | F      |

*In this truth table, the output of the 2-input AND gate is the same as the output of the 2-input OR gate. This is because when either A or B is true, the output of the AND gate is false and the output of the OR gate is true. When both A and B are*

*false, the output of the AND gate is false and the output of the OR gate is false. This shows that a 2-input AND produces an output that is the same as a 2-input OR gate.*

This issue is overcome by altering the wording of the prompt, instead of using "Show that", the prompt was rephrased as an explicit question using "Does a" as the opening phrase. That change flips ChatGPT to evaluate the truth of the statement which it does correctly:

*Does a 2 input AND gate produce an output that is the same as a 2 input OR Gate?*

*No, a 2-input AND gate produces an output that is the opposite of a 2-input OR gate. The output of a 2-input AND gate will be high (1) only if both inputs are high (1). The output of a 2-input OR gate will be high (1) if either or both inputs are high (1).*

The answer provided by ChatGPT is simply trying to fulfil the request posed to it, hence the answer does "show that" AND = OR even if that is completely incorrect. ChatGPT suffers from "Hallucinations" as do all LLMs (Bang, et al., 2023; Davis, 2023). These hallucinations can cause ChatGPT to make difficult to verify statements that seem plausible but do not stand up to scrutiny and are arbitrary in nature. The AND = OR examples would seem to be a combination of a hallucination and an eagerness to please the user.

### 3.4 Limitations

This work merely scratches the surface of what the limitations and strengths of ChatGPT for spreadsheet code generation. A much larger and more exhaustive study is required to confirm the conclusions of this work and to understand the scope in which ChatGPT can be accurate and where its hard limits in ability exist.

One could argue that giving ChatGPT queries that require deduction and inference with incomplete information is not a fair test of the technology and that no claims were made around such abilities by the vendors of the software. Whilst this is true, it is likely that humans using this technology will do so in an imperfect way and it may become a tool for problem solving issues or for bridging gaps in programming knowledge, hence ChatGPT and other LLM presents anew kind of spreadsheet user risk

Whilst the data in this paper has revealed some interesting subtleties around how prompts are worded, much more study is needed to fully understand these issues. The emerging discipline of "Prompt Engineering" will be an important methodology for leveraging the most value from ChatGPT (White, et al., 2023; Short & Short, 2023).

### 3.5 Future work

From these experiments, some areas of future work are identified:

The first would be a more exhaustive examination of the programming pitfalls and strengths of ChatGPT. This could take the form of further experiments into producing spreadsheet formulae with uncertainty in the prompts to find the limits of what is possible with the technology. The use of language in the prompt seems to be particularly significant to the responses generated by the software, some detailed examination of how ChatGPT responds to the use of different language phrasing could be beneficial to understanding how to best exploit the technology. This could establish the foundation of "Prompt Engineering" for spreadsheet applications based on some of the findings for general purpose programming (White, et al., 2023; Short & Short, 2023)

The second area of future work would be an empirical study of spreadsheet modellers who use ChatGPT to produce formulae. This study could examine the way in which the technology is being used by real spreadsheet modelers including their approach to generating, testing and verifying the output of ChatGPT. This goal of this work would be to understand best practice in spreadsheet modelling with ChatGPT and to understand the practical strengths and pitfalls of the technology.

The third area of future research could experimentally test the efficiency and efficacy of spreadsheet modellers using ChatGPT versus spreadsheet modellers not using ChatGPT. The aim of this work would be to understand the usefulness of the approach across different levels of experience and could be measured in terms of how accurate or otherwise each group is at generating spreadsheets for certain problems and how much time and effort is invested into producing the answer.

The final area of future work would be compare different AI based approaches to creating spreadsheets. For instance, contrasting LLMs with alternative approaches that leverage AI in different ways such as (Hofer, et al., 2017; Thorne, et al., 2013)

### 3.6 Conclusion

ChatGPT is a powerful resource that has great utility in certain activities, particularly for natural language generation. It also has some ability in code generation. Where the prompt is offered in complete detail, in these circumstances ChatGPT provided consistently correct code. However, as these experiments have show, if there is any uncertainty, inference or deduction needed from the prompt, ChatGPT has questionable ability to provide accurate code or reasoning. This opens a new front in spreadsheet risks, those that arise from the use of LLM generated spreadsheet formulae.

# REFERENCES

Bang, Y. et al., 2023. *A Multitask, Multilingual, Multimodal Evaluation of ChatGPT,* Pre-print Arxiv, Centre for Artificial Intelligence Research (CAiRE) The Hong Kong University of Science and Technology, 2302.04023.pdf (arxiv.org)

Borji, A., 2023. *A Categorical Archive of ChatGPT Failures,* Pre-print Arxiv, Quintic AI, 2302.03494.pdf (arxiv.org)

Chen, M. et al., 2021. *Evaluating Large Language Models Trained on Code,* Pre-print Arxiv, Evaluating Large Language Models Trained on Code (arxiv.org)

Davis, E., 2023. *Mathematics, word problems, common sense, and artificial intelligence,* Pre-print Arxiv, Department of Computer Science, University of New York, 2301.09723.pdf (arxiv.org)

Frieder, S. et al., 2023. *Mathematical Capabilities of ChatGPT,* Pre-print Arxiv, Universities of Oxford, TU Wien, Cambridge, Vienna and Princeton, 2301.13867.pdf (arxiv.org)

Hofer, B., Nica, I., Wotawa, F., 2017, *AI for Localizing Faults in Spreadsheets*. In: Yevtushenko, N., Cavalli, A., Yenigün, H. (eds) Testing Software and Systems. ICTSS 2017. Lecture Notes in Computer Science(), vol 10533. Springer, Cham. https://doi.org/10.1007/978-3-319-67549-7_5

Irons, R., 2003. *The Wall and The Ball: A study of domain referent spreadsheet errors,* Proceedings of the third annual conference of the European Spreadsheets Risks Interest Group (EuSpRIG), Dublin pp. 1-15, Microsoft Word - The Wall and The Ball _2_.doc (eusprig.org)

Jiang, Z., Xu, F., Araki, J. & Neubig, G., 2020. How Can We Know What Language Models Know? *Transactions of the Association for Computational Linguistics,* Volume 8, pp. 423-438.

Panko, R. & Halverson, R., 2001. An Experiment in Collaborative Spreadsheet Development. *Journal of the Association for Information Systems,* 2(4).

Panko, R. & Sprauge, R., 1998. Hitting the Wall: Errors in Developing and Code Inspecting a "Simple" Spreadsheet Model. *Decision Support Systems,* 22(4).

Rakovic, L., Sakal, M. & Vukovic, V., 2019. Improvement of Spreadsheet Quality through Reduction of End-User Overconfidence: Case Study. *Periodica Polytechnica Social and Management Sciences,* pp. 19-130.

Short, C. & Short, J., 2023. The artificially intelligent entrepreneur: ChatGPT, prompt engineering, and entrepreneurial rhetoric creation. *Journal of Business Venturing Insights,* Volume 19.

Teo, T. & Tan, M., 1997. *Quantitative and Qualitative Errors in Spreadsheet Development.* Maui, Hawaii, IEEE, pp. 149-155.

Thorne, S. Ball, D. & Lawson, Z., 2013, Reducing Error in Spreadsheets: Example Driven Modeling Versus Traditional Programming, *International Journal of Human–Computer Interaction*, 29:1, 40-53, DOI: 10.1080/10447318.2012.677744

Wang, S., Scells, H., Zuccon, G. & Koopman, B., 2023. *Can ChatGPT Write a Good Boolean Query for Systematic Review,* s.l.: Pre Print ArXiv.

White, J. et al., 2023. *A Prompt Pattern Catalog to Enhance Prompt Engineering with ChatGPT,* Pre-print Arxiv, Department of Computer Science Vanderbilt University, 2302.11382.pdf (arxiv.org)